# Downturn LGD:
# A More Conservative Approach
# for Economic Decline Periods

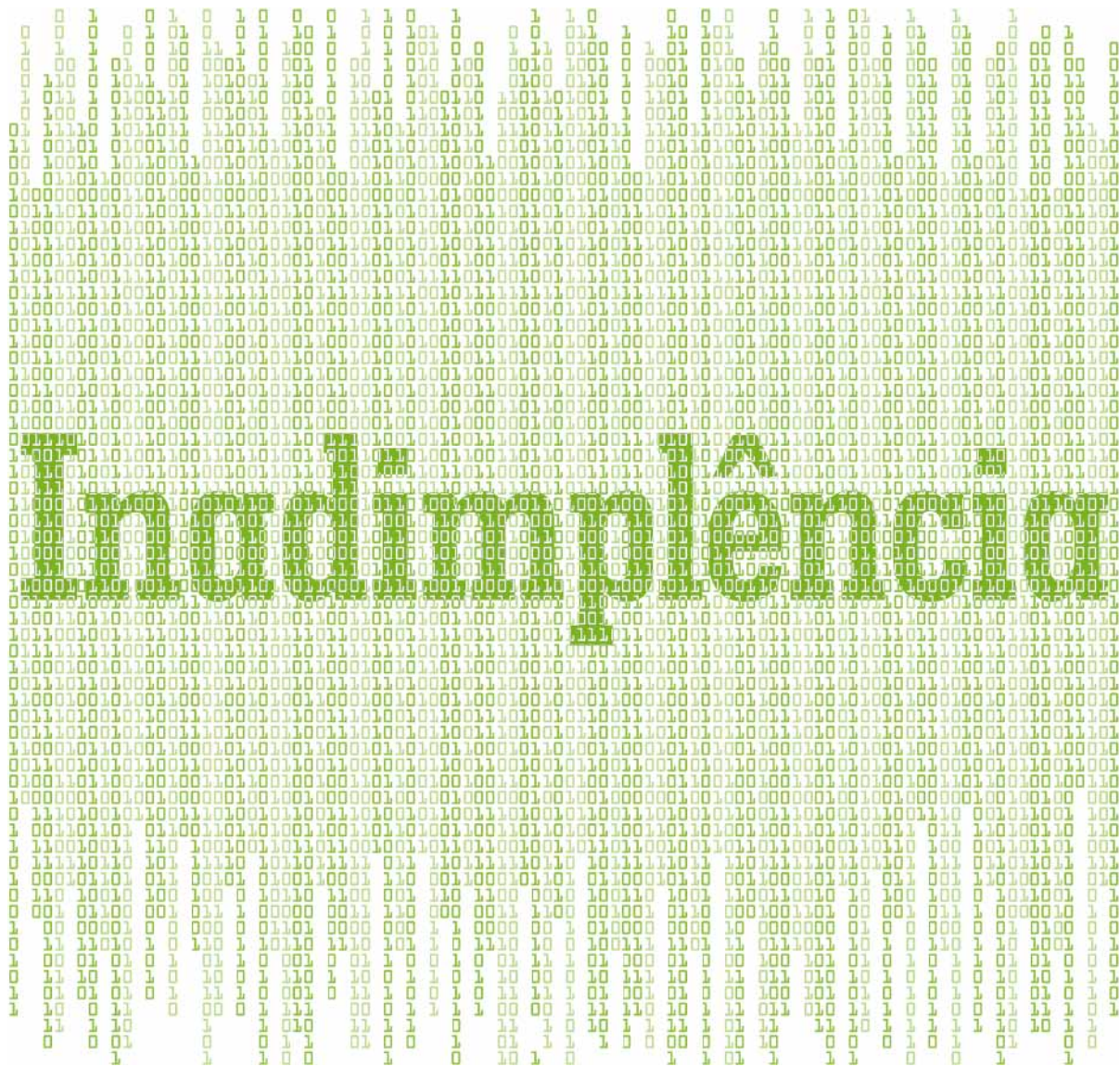

## Mauro Ribeiro de Oliveira Júnior
## Armando Chinelatto Neto

## Abstract

The purpose of this paper is to identify a relevant statistical correlation between rate of default (RD) and loss given default (LGD) in a major Brazilian financial institution's Retail Home Equity exposure rated using the IRB approach, so that we may find a causal relationship between the two risk parameters. Therefore, according to Central Bank of Brazil requirements, a methodology is applied to add conservatism to the estimation of the Loss Given Default (LGD) parameter at times of economic decline, reflected as increased rates of default.

## 1. Introduction

Financial institutions that wish to use IRB (Internal Ratings Based) approaches pursuant to the recommendations of the Basel Committee on Banking Supervision as provided in the document titled "International Convergence of Capital Measurement and Capital Standards: a Revised Framework", or simply Basel II, must calculate the following risk parameters: Probability of Default (PD), Loss Given Default (LGD), Exposure at Default (EAD) and Effective Maturity (M).

Central Bank of Brazil Communiqué No. 18.365 (2009) established the preliminary guidelines for using these approaches, and allows larger banks to use the advanced approach based on the internal rating of exposures according to their credit risk, that is, the Advanced IRB approach, to which the literature refers as A-IRB.

Central Bank of Brazil Circular Letter No. 3.581, dated March 8, 2012, addresses the standards governing the use of internal credit risk systems. In practice, Circular Letter No. 3.581 tropicalized the concepts of Basel II.

According to Carvalho and Santos (2009) the A-IRB approach requires the most sophisticated models and is therefore more complex than the other approaches that Basel II discusses.

Under A-IRB, Central Bank authorized institutions will be able to used their own estimates for Probability of Default (PD), Loss Given Default (LGD), Exposure at Default (EAD) and Effective Maturity (M), subject, at all times, to minimum criteria set forth by the supervising authority.

LGD corresponds to the percentage loss relative to total exposure at the time of default, and must be determined for every contract in the portfolio that is in default status.

According to article 75 of Circular Letter no. 3.581 and to paragraph 468 of Basel II, banks using the A-IRB method must estimate LGD parameters so that they cover a full economic cycle and are equal to or greater than the long-term weighted average of observed LGD percentages.

According to Communiqué No. 18.365 and Circular Letter No. 3.581, where the losses show cyclicality characteristics, the LGD parameter must reflect periods of adverse economic conditions, a practice that is referred to in the literature as "downturn LGD".

Indeed, at times of economic downturn, obligors in arrears may have increased difficulty honoring their obligations, leading to higher loss rates for banks' portfolios. In this case, the mean historic LGD, which may include a previous period of "economic normalcy", will not provide an accurate indicator of the future losses faced as an adverse period arises, and will, in fact, underestimate the portfolio's real loss potential.

It is during recessive periods that both retail and wholesale loan portfolios face their greatest rates of default: as the economic scenario deteriorates, it may compromise obligors' capacity to honor their debts.

Therefore, during economic downturns, a portfolio's loss rates may be greater than during normal periods, and the portfolio's downturn LGD must reflect this volatility.

Circular Letter No. 3.581 addressed this point explicitly, as it provides that LGD estimates must be more conservative in the presence of a relevant positive correlation between the frequency of default and the magnitude of LGD.

Given the above, calculating a downturn LGD is justified because, in almost every case, the LGD obtained from the historic average of recovery rates may not be independent from rates of default, requiring an added penalty for LGD estimates for the duration of economic downturn periods.

In light of the foregoing, this paper's research problem is to propose a downturn LGD for a Retail Home Equity portfolio that can reflect added conservatism at times when default rates are on the rise.

Section 2 analyzes an overview of the available literature on rates of recovery and rates of default. Section 3 shows the application of the methodology developed in the literature to calculation of a downturn LGD. Finally, Section 4 presents the paper's conclusions.

## 2. Literature Review

The academic studies reviewed show that economic adversity and periods of high default imply greater expected losses. It is therefore a mistake to regard LGD as independent from rates of default (RD) and use only the long-term average LGD to estimate the portfolio's future LGD.

The conservatism supervisors require in connection with LGD estimates, according to Altman and Ssabato (2005), is due to the fact that Pillar 1 capital requirements under Basel II are highly sensitive to the magnitude of LGD, particularly as concerns retail asset classes. Therefore, requiring the calculation of a more conservative – downturn – LGD is intended to make sure that capital requirements accurately reflect the capital needed to face unexpected losses arising from credit portfolio exposures over long periods of time.

Most of the academic studies on the dependence between the behavior of LGD and variables indicating economic downturns analyze data for debt issued by large businesses. For example, Altman et al. (2005) examine the rates of recovery of corporate obligations in 1982-2002 to conclude that macroeconomic variables do answer for a small part of the variation in rates of recovery.

Arguments for evidence of dependence between the recovery of defaulting exposures and variables associated with recessive economic conditions can be found in Frey (2009), where the author shows that, during recessions, the recovery of defaulted obligations is around 30% lower than during periods of economic growth. His study, however, also examines American corporate bonds, where a high correlation between default and recovery can be found.

Basel Committee 2005 document "Guidance on Paragraph 468 of the Framework Document" helps banks interpret paragraph 468 of Basel II as concerns the demand for downturn LGD.

The document describes the process to be followed to evaluate the potential effects of worsening economic conditions on recovery rates. According to the guidelines, the first step is to identify a historic period that can be characterized as a period of economic downturn.

One understanding that can be derived from the document is that, if the recovery rates observed during the periods of highest defaulting are lower than the average long-term rates of recovery, there is a potential for increased losses in periods of rising default. Therefore, failing to adopt a conservative LGD estimate for this period may underestimate the capital needed to cover unexpected losses.

We thus conclude that the appropriate approach to identify a period of economic decline must be based on tracking the historic observed rates of default and, as a result, periods in which historic observed rates of default are high will be then associated to a specific period of economic downturn for each credit portfolio.

Our review of the literature revealed a single article that uses data for retail assets: Sabato (2009). According to the author, it was the first study that proposed to analyze the ties between rates of recovery and an economic downturn for this asset class.

In the paper's results, the author showed a high and positive Pearson correlation coefficient of .77 between LGD and default rates. The correlation was obtained using data from a portfolio of products categorized, as per Basel II, as Other Retail Exposures - ORE.

On the other hand, in the same study, the author demonstrated the absence of significant correlation between variables in retail portfolios of products categorized as Qualified Revolving Exposures (-.84) and in Exposures to Small and Medium Enterprises (-.12), suggesting that these two asset classes do not require calculation of a downturn LGD.

Identification of a statistical connection between two variables may not establish a casual tie between them. In this paper, we use a method supported by the literature and known as the Granger Causality Test, which allows finding a causal link (or temporal precedence) between any two variables, in the sense that variable X Grange-causes variable Y if the observed X in the present or past helps predict the future values of Y for a given horizon (GRANGER, 1969).

Within the context of this paper, if the granger RD causes LGD, then changes in RD must precede changes in LGD over time.

However, the two variables must be stationary, as the stationary condition is crucial to application of the Granger causality test (GUJARATI, 2004).

Data constraints are a significant challenge in estimating LGD parameters in general and downturn LGD parameters in particular. Furthermore, there is no certain, supervisor-provided, methodology in terms of what methods are appropriate for conservative estimates of LGD under economic downturn conditions.

Therefore, according to the method proposed in Sabato (2009), we regard an economic downturn as a minimum period of six months during which the observed rate of default is consistently higher than the long-term historic rate plus one standard deviation.

The extent to which potential dependence between the rates of default and LGD will imply increased conservatism in A-IRB models may vary significantly from one financial institution to another, given that the methodologies take internal databases into consideration for the purposes risk-parameter calculation.

Therefore, the purpose of this study – that is, to calculate downturn LGD based on historic data for the Home Equity Retail sub-class, is to make sure that LGD includes future predictions of loss rates relative to the risks that increased defaulting may create.

## 3. Methodology

The method proposed above to calculate downturn LGD consists of identifying the period of economic downturn and then calculating how much, in percentage terms, loss given default increases relative to periods of lower-than-average rates of default.

Illustrating the method with data from Other Retail Exposures, as in Sabato (2009), the economic downturn period showed an average 17% increase in LGD for the category relative to the period in its entirety. This factor must be added to the portfolio's LGD to increase the variable's conservatism for the period of deteriorating economic conditions, that is, when the frequency of default starts rising to above-average levels.

This paper adopted the downturn LGD development model proposed in Sabato (2009) and applied the technique to Home Equity Retail exposures of a major financial institution active in the domestic financial market.

The LGD database consisted of

monthly observed LGD data for the period from January 2008 to November 2011 (see graphic 1). The rate of default database was obtained from the Central Bank of Brazil Website and comprehend the same period, that is, January 2008 to November 2011.

The .71 correlation between loss given default and the rate of default is significant for the purposes of the study.

■ Graphic 1

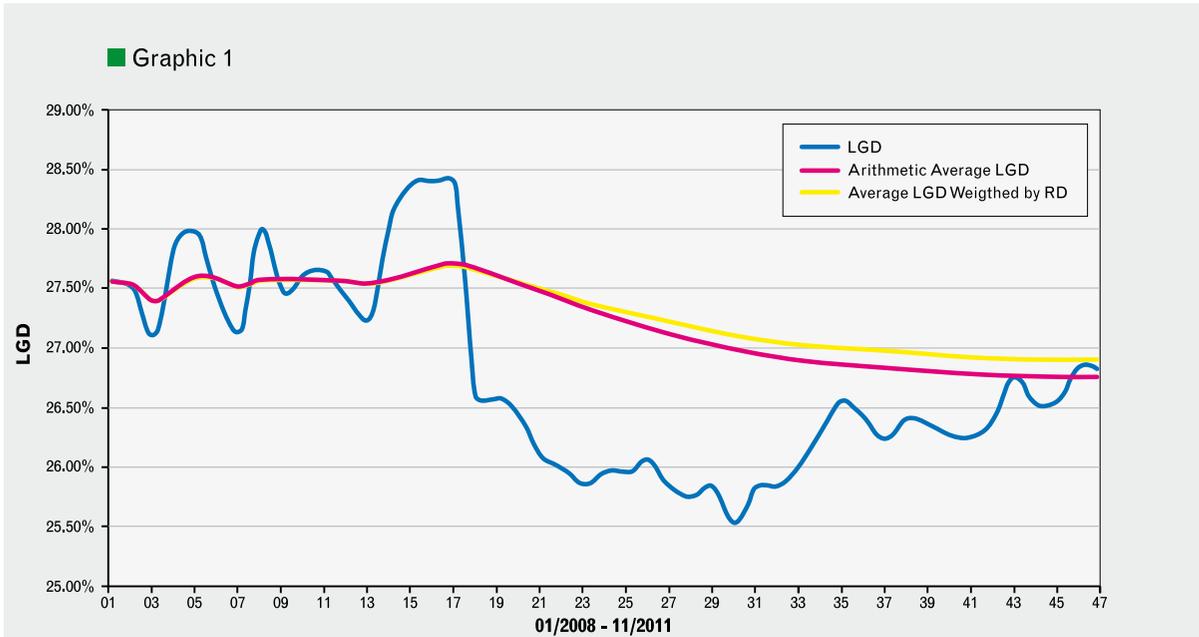

■ Graphic 2

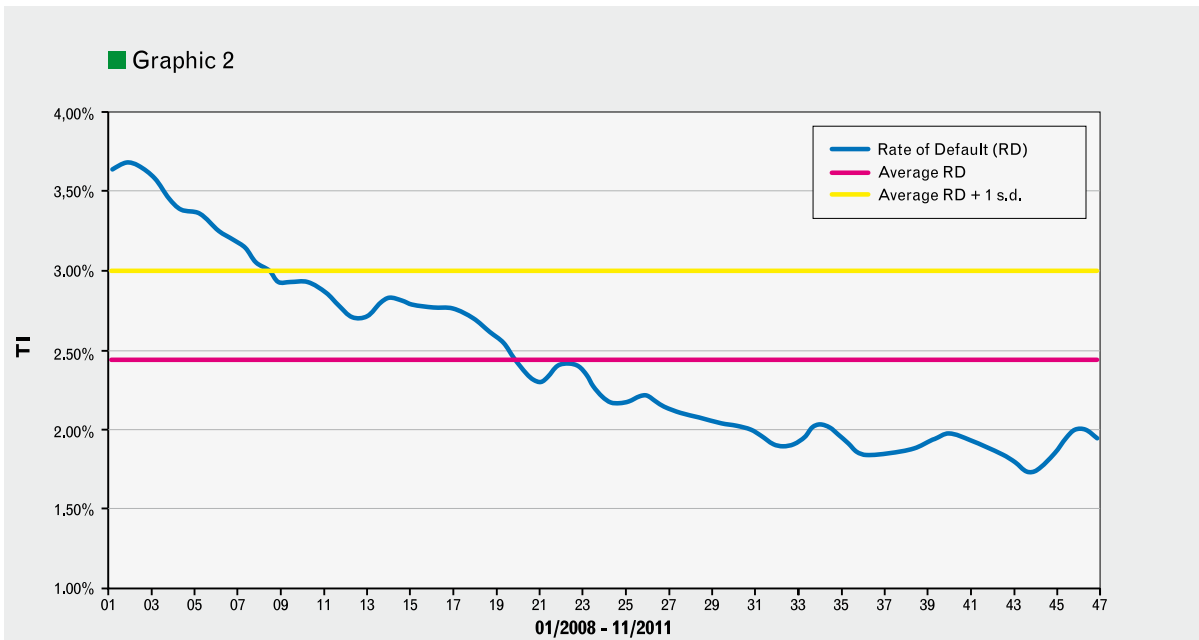

### 3.1. Granger Causality Test for RD and LGD

Although a strong Pearson correlation exists between the variables involved in our study, this alone does not mean that a causal tie is present, that is, that an increase in RD in a certain period does affect the increase in LGD in a subsequent period.

The Granger Causality test is a VAR model, that is, a Vector Autoregressive model that examines linear relationships between each variable and the lagged values of itself and every other variable. Such models consider the presence of interdependence between variables and may assess the dynamic impact of random disturbances on the variables system, which makes them particularly useful and efficient in predicting the future behavior of interrelated time series (CAIADO, 2002).

The stationary condition of the RD and LGD series is crucial to application of the Granger Causality Test method for the purposes of capturing causal ties between the variables. It is important to point out that if a series displays a unit-root, then it is non-stationary.

The analyses show that LGD does have a unit-root, that is, is non-stationary. On the other hand, RD is stationary, as it lacks a unit-root. However, given the constraints of the data period, we decided not to develop vector models based on the cointegration of the variables.

Assuming LGD to be stationary, the results we found using Eviews 5 and adopting a 5% significance level for each of the two tests, is that LGD causes RD (for 5 lags).

This means that today's changes in LGD will affect RD in 5 periods, that is, today's RD is impacted by the LGD from 5 periods ago, at a 5% significance level.

### 3.2. Economic Downturn Period

As noted, the approach to identifying a period of economic downturn will be based on historic observed rates of default.

We therefore considered a period of economic downturn to be a minimum period of six months during which the observed rated of default is consistently higher than the historic long-term rate plus one standard deviation.

Give the information from our database, the period from January 1208 to August 2008 show rates of default in excess of 2.99%, which is the historic average plus one standard deviation. Therefore, this period will be regarded as a period reflecting adverse economic conditions.

### 3.3. Expected Average Increase in LGD

More conservatively, the average expected increase in LGD may be calculated as the relationship between the downturn period's weighted average LGD plus one standard deviation and the average LGD of the period of lowest rates of default minus one standard deviation.

Our result is an LGD that is 7.73% higher during recessive periods than in periods of low defaulting. This implies an LGD that is 2 percentage points higher.

In this case, the formula for the selected portfolio would be:

$$dLGD1 = 0.02 + ELGD \text{ (graphic 3)}$$

On the other hand, according to a slightly less conservative view, the average expected increase in LGD may be calculated as the relationship between the downturn weighted average LGD plus one standard deviation and the average LGD for the entire period minus one standard deviation.

In this case, the result is an LGD that is 7% higher during recessive periods than for the portfolio's entire period. This implies an LGD that is 1.83 percentage point higher.

In this case, the formula for the selected portfolio would be:

$$dLGD2 = 0.0183 + ELGD \text{ (graphic 3)}$$

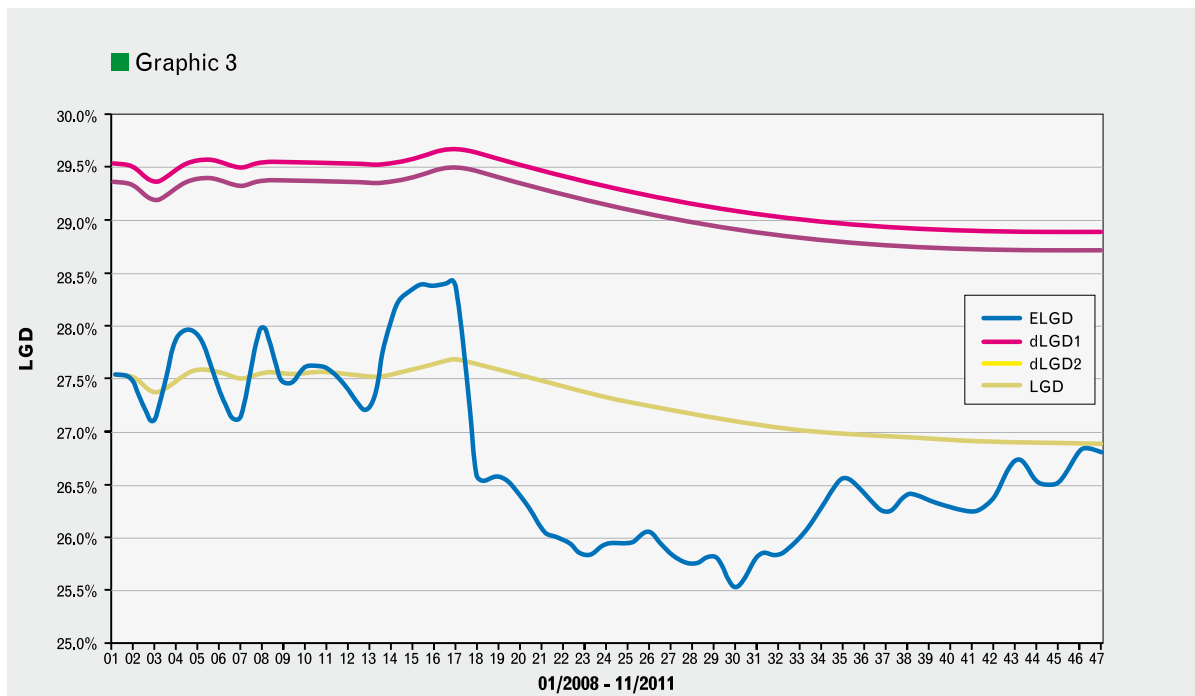

■ Graphic 3

01/2008 - 11/2011

## 4. Conclusion

This paper investigated a methodology that adds conservatism to LGD estimates at times when the beginning of an economic downturn is perceived. We developed a methodology that has been established for retail portfolios, but had never been applied to a Home Equity Retail portfolio.

Despite the high statistical correlation between RD and LGD, we cannot, based on the statistical tools available, claim that the period marked by an economic downturn, that is, the period in which rates of default follow a rising path, may have a future impact on increased economic losses arising from defaults.

However, despite not having found that RD Granger-causes LGD, LGD must even so, as per the Central Bank's recommendations, be more conservative in periods when default begins to rise. This paper attempted to determine how much more conservative we should be by applying an added penalty to estimations of LGD.

We therefore tried to meet the Central Bank of Brazil's requirements for the LGD estimates used in A-IRB. For future and supplementary studies, we suggest calculating more conservative metrics for portfolios where a significant correlation and time causality exist between rates of recovery and collateral value, since, if such a relationship can be found, the deterioration of collateral values will bring about a potential increase in default-related losses and, consequently, in LGD.

**Autores**

### Mauro Ribeiro de Oliveira Júnior

Bachelor of Mathematics, Universidade Federal de São Carlos, Master of Mathematics, Universidade Estadual de Campinas, MBA in Risk Management, Fundação Instituto de Pesquisas Contábeis, Atuarias e Financeiras – FIPE-CAFI. Currently a Doctor of Statistics candidate at Universidade Federal de São Carlos and employed at a domestic retail bank. Contact the author: *mauroexatas@gmail.com*

### Armando Chinelatto Neto

Bachelor of Economics and Master and Doctor of Applied Economics, Universidade Federal de Viçosa – UFV. MBA in Risk Management, Fundação Instituto de Pesquisas Contábeis, Atuarias e Financeiras – FIPECAFI. Consultant to the CEO at a domestic retail bank. Contact the author: *armandochinelattoneto@gmail.com*